\documentclass[twoside]{article}
\usepackage{fleqn,espcrc2}

\usepackage{graphicx}

\newcommand{\AmS}{{\protect\the\textfont2
  A\kern-.1667em\lower.5ex\hbox{M}\kern-.125emS}}

\title{The standard model and parity conservation}

\author{She-Sheng Xue\address[MCSD]{ICRA, INFN  and
Physics Department, University of Rome ``La Sapienza", 00185 Rome, Italy
}%
}
      
\begin{document}

\begin{abstract}
On the basis of previous work on chiral gauged fermions on a lattice, we discuss the lattice-regularization of the standard model by introducing two Weyl fields interacting with quarks and leptons. These 
interactions form massive bound states to gauge-invariantly decouple doublers at 
high energies and these bound states dissolve into their constituents at low energies. No any hard 
spontaneous symmetry breakings occur at the lattice scale $\pi/a$. As a consequence, the 
gauge symmetries of the standard model are realized by both massive vectorlike
spectra at high energies and massless chiral spectra at low energies. Such a scenario is consistent with the gauge-anomaly cancelation, flavor-singlet anomaly and Witten's anomaly. These studies predict that the parity symmetry must be restored at high energies.

\vspace{1pc}
\end{abstract}

\maketitle

\section{Introduction}

The parity-violating feature at low-energies is strongly phenomenologically supported. Based on this feature, the extremely successful standard model for particle physics is constructed in the form of a renormalizable quantum field theory with chiral gauge symmetries. While, the very-small-scale structure of the space-time, the arena of physical reality, can exhibit rather complex structure of a space-time ``string" or ``foam", instead of a simple space-time point. As the consequence of these fundamental constituents of the space-time, the physical space-time gets endowed with a fundamental length that provides a neutral regularization for a fundamental theory, as a result, fundamental theory must be finite. However, with very generic axioms, the ``no-go" theorem \cite{nn81} demonstrates that the quantum field theories with parity-violating gauge symmetries, as the standard model (SM), cannot be consistently regularized on the lattice, i.e., the discrete space-time. The vectorlike fermion spectrum and parity-conservation gauge coupling are unavoidable at the lattice scale $\pi/a$, as a consequence, the gauge anomaly and flavor singlet anomaly are not correctly produced. This paradox may imply a new physics beyond SM at low energies. This has been, for a long time, our basic point to find a resolution to this paradox. 
Searching for a chiral gauge symmetric approach to properly regularize SM on the lattice has been greatly challenging to theoretical particle physicists for the last two decades. 

One of the classes of approaches is the modeling 
by appropriately introducing local interactions\cite{ep}. However, the phenomenon of hard spontaneous symmetry breakings in the intermediate coupling and the argument of anomaly-cancelation within vectorlike spectra in the strong coupling phase prevent such modelings from having a scaling region for low-energy chiral gauged fermions. It was then a general belief that it seems impossible to formulate a theory of exact chiral gauge symmetries on the lattice. Nevertheless, in refs.\cite{xue97}, we advocated a model with peculiar 10-dimension four-fermion interactions on the lattice and a plausible scaling region at low energies. The dynamics of realizing chiral gauged fermions was studied\cite{xue97l,xue00d}. In ref.\cite{xue00n}, we computed and discussed the gauge anomalies and flavor singlet anomalies. In ref.\cite{xue01}, based on a simple chiral-fermion model (11112) on the 1+1 dimensional lattice, we gave an exact illustration of the dynamics realizing chiral gauge theories in the low-energy scaling region, which were discussed in previous papers\cite{xue97,xue00n}. In this paper, with two additional Weyl fermions: sterile neutrino $\chi_L$ and spectators $\chi_R=\nu_R$, we analogously introduce strong four-fermion interactions for leptons and quarks, we present a regularization of SM on the lattice and try to show the dynamics that we studied in the previous papers\cite{xue97,xue00n} can realize SM at low energies. We obtain a consistent scenario and predict that the parity conservation must be restored at high energies.

\section{The model}

We consider the first generation of SM, whose fermion content consists of the left-handed doublets $\psi_L=(\nu_L,e_L),(u_L,d_L)$ and  right-handed singlets $\psi_R=e_R,u_R,d_R$. The Higgs sector is disregarded and all fermions are massless. At high energies of the lattice scale, the naive lattice-regularized action for SM is given by
\begin{equation}
S_\circ \!=\!{1\over 2a}\sum_x\Big(\bar\psi_L(x) D^L_\mu\cdot\gamma_\mu\psi_L(x)+(L\rightarrow R)\Big),
\label{free}
\end{equation}
where $D_\mu^{L,R}$ are ($\delta_{x,x\pm 1}\psi(x)=\psi(x\pm a_\mu)$),
\begin{equation} 
D_\mu^{L,R}=([U_\mu(x)]^{L,R}\delta_{x,x+1}
-[U_\mu^\dagger(x)]^{L,R}\delta_{x,x-1}),
\label{kinetic} 
\end{equation} 
where $U_\mu(x)$ is the gauge fields of the $SU_c(3)\otimes SU_L(2)\otimes U_Y(1)$ group of SM.  

We introduce two neutral Weyl fermions $\chi_L,\chi_R$. At the lattice scale, the $\chi_R$ couples to the left-handed doublets $\psi_L$ and $\chi_L$ couples to the right-handed singlets $\psi_R$ by following 10-dimension four-fermion interactions,
\begin{eqnarray}
S^L_i\!&=&\!g\bar\psi_L(x)\cdot\left[\Delta\chi_R(x)\right]
\left[\Delta\bar\chi_R(x)\right]\cdot\psi_L(x),
\label{hil}\\
S^R_i\!&=&\!g\bar\psi_R(x)\cdot\left[\Delta\chi_L(x)\right]
\left[\Delta\bar\chi_L(x)\right]\cdot\psi_R(x),
\label{hir}
\end{eqnarray}
where the coupling $g$ has dimension
$[a^{-2}]$ and the operator $\Delta$ is defined by,
\begin{equation}
\Delta\chi(x)\equiv\sum_\mu
\Big[ \chi(x+a_\mu)+\chi(x-a_\mu)-2\chi(x)\Big],
\label{wisf0}
\end{equation}
and its Fourier transformation
\begin{equation}
\Delta(p)=\sum_\mu\left(\cos(p_\mu)-1\right),
\label{wisf}
\end{equation}
where the dimensionless momentum $p_\mu=\tilde p_\mu a$. 
Eq.(\ref{wisf}) indicates that large momentum states of $\chi_R(\chi_L)$ strongly couple to $\psi_L(\psi_R)$, while small momentum states of $\chi_R(\chi_L)$ weakly couple to $\psi_L(\psi_R)$.

The interactions (\ref{hil},\ref{hir}) preserve the gauge symmetries of SM and the global symmetries $U_{B,L}(1)$ for the baryon- and lepton-numbers at the lattice scale. For the second and third fermion generations, eq.(\ref{free}-\ref{hir}) are the same respectively for an exact flavor symmetry at the lattice scale. We can also consider the interaction vertex of t'Hooft type\cite{mc}.

Considering the case of the strong coupling $ga^2\gg 1$ phase, we rescale the fermion fields $\psi\rightarrow g^{1\over4}\psi$ to be dimensionless 
and rewrite,
\begin{eqnarray}
S&=&{1\over 2ag^{1\over2}}\sum_x\Big(\bar\psi_L(x) D^L_\mu\cdot\gamma_\mu\psi_L(x)
+\cdot\cdot\cdot\Big)\nonumber\\
&+& \sum_x S^L_i(x)+ (L\rightarrow R),
\label{action}
\end{eqnarray}
where the coupling $g$ in $S^{L,R}_i(x)$ is rescaled away and ``$\cdot\cdot\cdot$" stands for the kinetic terms for $\chi_L$. 

\section{No hard spontaneous breaking}

In addition to the gauge symmetry and global symmetries $U_{B,L}(1)$ of SM, this lattice-regularized action (\ref{action}) possesses the exact shift-symmetries of $\chi_R$ and $\chi_L$\cite{gp}. It was demonstrated\cite{xue97} that in the region $ga^2\gg 1$ there is no any spontaneous breakings of the shift-symmetries via  
\begin{equation}
\langle\bar\psi_L\chi_R\rangle\not=0\hskip0.5cm {\rm and}\hskip0.5cm \langle\bar\chi_L\psi_R\rangle\not=0,
\label{eve}
\end{equation}
by interactions (\ref{hil},\ref{hir}).
These shift-symmetries protect the right-handed doublets $\psi_L$ and left-handed singlets $\psi_R$ from coupling each other and guarantee the spectators $\chi_R,\chi_L$ decoupled from gauge fields and other fields at low energies. This is very important to obtain the low-energy spectra, gauge couplings and correct anomalies.

\section{Three-fermion states}

It was demonstrated\cite{xue97} that for the strong coupling region $ga^2\gg 1$ at high energies, $\Delta(p)\not=0, p_\mu\not=0$ where vectorlike doublers are, the following three-fermion states $\Psi_R$ and $\Psi_L$ are formed, 
\begin{equation}
\Psi_R\equiv {1\over2a}(\bar\chi_R\cdot\psi_L)\chi_R,\hskip 0.3cm \psi_L=\left(\matrix{\nu\cr e}\right)_L,\left(\matrix{u\cr d}\right)_L
\label{l3}
\end{equation}
for the left-handed doublets $\psi_L$ and
\begin{equation}
\Psi_L\equiv {1\over2a}(\bar\chi_L\cdot \psi_R)\chi_L,\hskip 0.5cm \psi_R=e_R, u_R, d_R
\label{r3}
\end{equation}
for the right-handed singlets $\psi_R$. The three-fermion state $\Psi_R(\Psi_L)$ has the same quantum numbers as $\psi_L(\psi_R)$ but opposite chirality. The massive vectorlike (Dirac) 
fermions
\begin{equation}
(\psi_L, \Psi_R)\hskip0.5cm {\rm and}\hskip0.5cm(\psi_R, \Psi_L),
\label{r4}
\end{equation}
are formed consistently with the gauge symmetry $SU_c(3)\otimes SU_L(2)\otimes U_Y(1)$. 15 doublers at $p_\mu\not=0$ are thus gauge invariant decoupled from the low-energy spectrum. Non-perturbative $(O(1/a))$ and non-smooth $(O(a))$ variations of longitudinal gauge fields are eliminated.

\section{Dissolving at an intermediate scale}

In ref.\cite{xue97l,xue01}, we show that with the fixed value of the strong coupling $ga$, the effective coupling $\tilde ga(p)$ is weak at low energies $\Delta(p)\simeq 0, p_\mu\simeq 0$, there is an intermediate scale $\epsilon$
\begin{equation}
\tilde v \ll\epsilon \ll {\pi\over a},\hskip0.5cm \tilde v\sim 250GeV,
\label{epsilon}
\end{equation}
where $\tilde v$ is the electroweak scale.
At this intermediate scale $\epsilon$, the three-fermion states 
$\Psi_L$ and $\Psi_R$ must dissolve into the corresponding three-fermion cuts
\begin{equation}
{\cal C}[\Psi_L]\hskip0.5cm {\rm and}\hskip0.5cm{\cal C}[\Psi_R]
\label{cut}
\end{equation}
as binding energies go to zero and the residues of the propagators for $\Psi_{R,L}$ go to zero when $p\rightarrow \epsilon a\sim 0$.  These three-fermion cuts ${\cal C}[\Psi_L]$ and ${\cal C}[\Psi_R]$ are the virtual states processing the same quantum numbers as $\Psi_L$ and $\Psi_R$, such a dissolving process is thus gauge invariant. The low-energy spectra (below the threshold $\epsilon$) are those $\psi_L$ and $\psi_R$ in SM, in addition two free particles $\chi_R$ and $\chi_L$.
In Eq.(\ref{epsilon}), the value of the scale $\epsilon$ depends on the value of the coupling $ga^2$.
The electroweak scale $\tilde v$ is the soft spontaneous symmetry breaking $\langle\bar\psi_L\psi_R\rangle$ that can be introduced by the induced 6-dimension four-fermion interactions $\bar\psi_L\psi_R\bar\psi_R\psi_L$ at the intermediate scale $\epsilon$, rather than the lattice scale\cite{xue97}.

\section{Gauge couplings}

Above the threshold $\epsilon$, The 1PI vertex functions for gauge couplings are vectorlike and given by two parts\cite{xue97l,xue00n}, one is with and another without external truncated three-fermion states (\ref{l3},\ref{r3}), e.g.,
\begin{eqnarray}
&&\Lambda_\mu(p,p')=\Lambda_{\mu LL}(p,p')+\Lambda_{\mu LR}(p,p',\Psi_R)\nonumber\\
&&+\Lambda_{\mu RL}(p,p',\Psi_R)+\Lambda_{\mu RR}(p,p',\Psi_R).
\label{vdirac}
\end{eqnarray}
These 1PI vertex functions obey Ward identities of
the gauge symmetries of SM at the lattice scale, since the gauge symmetries are
realized by the vectorlike and massive spectrum (\ref{r4}).
Below the threshold $\epsilon$, the parts with external truncated three-fermion states in (\ref{vdirac}) vanish and these 1PI vertex functions of gauge couplings reduce to the counterparts of SM at low energies up to some local counterterms\cite{xue97l,xue00n}.

\section{Anomalies}
  
We turn to discuss the gauge anomaly and the $B-L$ number violation\cite{xue97l,xue00n}. Since the high-energy spectra above the threshold $\epsilon$ are vectorlike consistently with the gauge symmetry of SM, these high-energy spectra do not have any contributions to the gauge anomaly. The fermionic modes that contribute to the gauge anomaly are those living below the threshold $\epsilon$. For computing the gauge anomalies up to some local counterterms, we can adopt a continuous regularization, for instance Pauli-Villars
one, at the scale $\epsilon$ for perturbative $O(a)$ and smooth $O(1/a)$ gauge fields, and the result as well-known is proportional to,
\begin{equation}
\sum_{\rm fermions}{\rm tr}(T^aT^aY),
\label{anomaly}
\end{equation}
where $T^a$ is some component of the weak isospin and $Y$ denotes the weak hypercharge. Eq.(\ref{anomaly}) vanishes for the fermion content of SM that is anomaly-free. This is an important self-consistency for (i) our gauge invariant 
action and dynamics to have the low-energy fermion spectrum of SM; (ii) introducing local counterterms to eliminate local and explicit breakings of gauge symmetries at the scale $\epsilon$, which are introduced by continuous regularizations.

To compute the $B-L$ number violation, the flavor-singlet anomaly due to the $SU_L(2)$ instanton effect, we can adopt the approach of computing the mixing anomalies\cite{preskill91,xue00n}. In this case, the ``mixing'' gauge group is
$SU_L(2)\otimes U_B(1)\otimes U_L(1)$, where $U_{B,L}(1)$ respectively associate to the flavor-singlet currents $j^B_\mu$ and $j^L_\mu$ of the quarks and leptons. In such an approach, the computation is the same as that for the gauge anomaly (\ref{anomaly}) and we just need to replace the hypercharge $Y$ in (\ref{anomaly}) by one, the generator of $U_{B,L}(1)$. As discussed in ref.\cite{xue00n}, due to completely decoupling of $\chi_{R,L}$ from gauge bosons, anomalous flavor-singlet currents of quarks and leptons are obtained up to local counterterms,    
\begin{equation}
\partial^\sigma j^{B,L}_\sigma = {i\over32\pi^2}\tilde F^{\mu\nu}F_{\mu\nu}.
\label{bl}
\end{equation}
These results lead to the $B-L$ number conservation and $B+L$ number violation,
\begin{equation} 
\partial^\sigma (j^B_\sigma+j^L_\sigma)={i\over16\pi^2}\tilde F^{\mu\nu}F_{\mu\nu}.
\label{b-l}
\end{equation}

The $SU_L(2)$ global anomaly of the Witten type appears for odd numbers of left-handed $SU_L(2)$ doublets. We do not find any inconsistency in our model, since there are two (even numbers) left-handed $SU_L(2)$ doublets $\psi_L$ for leptons and quarks at low energies.

\section{Some remarks and predictions}

The final scenario appears to be: (i) vectorlike fermion spectra (\ref{r3},\ref{r4}) and gauge couplings (\ref{vdirac}) at high energies; (ii) chiral fermion spectra and parity-violating gauge coupling at low energies.  
Such a scenario implies an important prediction that the parity-violating phenomenon is only at low energies and parity conservation must be restored at high-energies. Whether this is true can be experimentally tested by measuring the left-right asymmetry,
\begin{equation}
A_{LR}={\sigma_L-\sigma_R\over\sigma_L+\sigma_R}
\label{lras}
\end{equation}
at high energies $\sim\epsilon$ above the electroweak scale, where $\sigma_L(\sigma_R)$ denotes the cross section for an incident left-handed (right-handed) polarized electrons. $A_{LR}$ is related to gauge couplings to fermions.

Based on such a scenario, we had made some preliminary studies of phenomenological aspects on the generation of fermion masses, fine-tuning problem and masses and mixing angles of the quark and lepton sectors\cite{neu00}. The Flavor physics, neutrino physics and any experimental phenomena deviating from that of SM are interesting subjects to study within this scenario.


\begin{thebibliography}{9}

\bibitem{nn81}
H.B.~Nielsen and M.~Ninomiya, Nucl.~Phys.~B185 (1981) 20, {\it
ibid.}  B193 (1981) 173, Phys.~Lett. B105 (1981) 219.

\bibitem{ep}
E.~Eichten and J.~Preskill, Nucl.~ Phys. B268 (1986) 179.\\
J.~Smit, Acta Physica Polonica B17 (1986) 531;
P.D.V.~Swift, Phys.~Lett. B145 (1984) 256;\\
S.~Aoki, I-Hsiu Lee and S.-S.~Xue, Phys.~Lett. B229 (1989) 79.\\
G.~Preparata and S.-S.~Xue, Phys.~Lett. B264 (1991) 35.\\
I.~Montvay, Phys.~Lett. B199
(1987) 282; Nucl. Phys. B29 (Proc.~Suppl.)
(1992) 159, {\it ibid} 30B (1993) 621 
and references therein.\\
M.F.L.~Golterman, D.N.~Petcher and E.~Rivas, Nucl.\ Phys. B395 
(1993) 597.\\
D.N.~Petcher, Nucl.~Phys. (Proc.~Suppl.) B30 
(1993) 52, references there in.

\bibitem{xue97}
S.-S.~Xue, Phys.~Lett. B381 (1996) 277 and Nucl.~Phys. B486 (1997) 282.

\bibitem{xue97l}
S.-S.~Xue, Phys.~Lett. B408 (1997) 299; 
Nucl.~Phys. (Proc.~Suppl.) B53 (1997) 668. 

\bibitem{xue00d}
S.-S.~Xue, Phys.~Rev.~D61 (2000) 054502. 
 
\bibitem{xue00n}
S.-S.~Xue, Nucl.~Phys.~B580 (2000) 365 and references there in.

\bibitem{xue01}
S.-S.~Xue, hep-lat0006024.

\bibitem{mc}
M.~Creutz, C.~Rebbi, M.~Tytgat and S.-S.~Xue,
Phys.~Lett. B402 (1997) 341;\\
M.~Creutz,  Nucl.Phys.Proc.Suppl. 63 (1998) 599; S.-S.~Xue, {\it ibid} 63 (1998) 596.

\bibitem{gp}
M.F.L.~Golterman, D.N.~Petcher,  Phys.~Lett. B225 
(1989) 159. 

\bibitem{preskill91}
J.~Preskill,  Ann.~Phys. 210 (1991) 323.

\bibitem{neu00}
S.-S.~Xue, Phys.~Lett.~B 377 (1996) 124; {\it ibid} B 398 (1997) 177;
Mod.~Phys.~Lett.~A Vol.~14 (1999) 2701; {\it ibid} Vol.~15 (2000) 1089.


\end{thebibliography}
\end{document}